# Origin of life from a maker's perspective – focus on protocellular compartments in bottom-up synthetic biology

Authors: **Ivan Ivanov**[1,2], **Stoyan K. Smoukov**[3], **Ehsan Nourafkan**[3], **Katharina Landfester**[4], **Petra Schwille**[5]

[1]Max Planck Institute for Dynamics of Complex Technical Systems, Sandtorstraße 1, 39106, Magdeburg, Germany.
[2]Universitat Politècnica de Catalunya, Rambla Sant Nebridi 22, TR14 08222 Terrassa, Barcelona, Spain.
[3]Active and Intelligent Materials Lab, SEMS, Queen Mary University of London, London E1 4EN, UK.
[4]Max Planck Institute for Polymer Research, Ackermannweg 10, 55128, Mainz, Germany.
[5]Max Planck Institute of Biochemistry, Am Klopferspitz 18, 82152, Martinsried, Germany.
Correspondence: ivan.ivanov@upc.edu, s.smoukov@qmul.ac.uk, landfester@mpip-mainz.mpg.de, schwille@biochem.mpg.de

Abstract:

The origin of life is shrouded in mystery, with few surviving clues, obscured by evolutionary competition. Previous reviews have touched on the complementary approaches of top-down and bottom-up synthetic biology to augment our understanding of living systems. Here we point out the synergies between these fields, especially between bottom-up synthetic biology and origin of life research. We explore recent progress made in artificial cell compartmentation in line with the crowded cell, its metabolism, as well as cycles of growth and division, and how those efforts are starting to be combined. Though the complexity of current life is among its most striking characteristics, none of life's essential features require it, and they are unlikely to have emerged thus complex from the beginning. Rather than recovering the one true origin lost in time, current research converges towards reproducing the emergence of minimal life, by teasing out how complexity and evolution may arise from a set of essential components.

The question how life originated is difficult both because direct evidence is long gone and because modern definitions of life reveal it is a grayscale continuum phenomenon, not a binary yes/no test. One humbling yet motivating point of agreement is that life is a complex phenomenon that is not sufficiently understood and not yet possible to build in the lab – despite the Miller-Urey experiment[1,2] and the technological advancements in combinatorial chemistry,[3,4] fast-forwarding billions of years of stochastic conditions and interactions on a planetary scale. In an effort to strip down the skeleton of life, top-down synthetic biology approaches have modified living cells by, e.g. paring genomes down to a set of 473 genes capable of supporting replication, though the functions of a third of those is still unknown.[5] Bottom-up synthetic biology, in contrast, starts with few components, performing increasingly complex functions, with the aim of creating life.[6] Furthermore, it has the ability to take under its umbrella research from a number of scientific fields such as colloidal science, biochemistry, biophysics, genetic engineering, active matter, etc. and can be considered as a conceptual offspring of abiogenesis research. Thus, it has inherited many of the tools and aims of the latter, and the terms proto-, synthetic, artificial, and minimal cell are often used interchangeably. However, bottom-up synthetic biology is not burdened by the command of the evolutionary timeline – its molecular toolbox ranges from simplistic presumable predecessors to highly evolved biological machinery. It takes advantage of the newest developments in molecular biology and relies rather on controlled production, than on environmental cycles and chance. In addition, it is not limited to natural building blocks and architectures but augmented by the power of synthetic chemistry and subjected to modular reorganization at will, with the overarching aim of reproducing essential biological features with a minimal set of parts.

Similar two-sided approaches in the exploration of life's emergence and evolution are attempted all the way from defining the chemistry of proto-life, to the higher functions above mostly considered the domain of biology, with a palpable tension. On one hand there is a desire to "recognize life when we see it" often implying not only DNA but most of its other modern properties, as if in the Greek myth seeing Aphrodite (Venus) emerging fully formed from the ocean foam [7]. On the other, it is a self-evident postulate that the intricate constituents and complexity of modern biology did not emerge at once in that prebiotic time. Therefore, it has been necessary to separate and conceptualize the component functions by which we might recognize life more abstractly, including not only the above movement, metabolism, communication, reproduction, and evolution. Paring down the origin of more advanced cell properties, such as compartmentation, current metabolic cycles, or the present DNA-RNA-protein machinery will bring additional fundamental insights but may also trigger new applications. For example, minimal chemical systems have been proposed to randomly generate predecessors and plausible molecular pathways to the molecules of life [4]. But such systems also have the highly sought application of designing the synthesis for any molecule. Equally simple "biological" ensembles are put together from few components to mimic ever larger collections of features of life. In both chemical and biological systems (we note the increasingly arbitrary discrimination) scientists aim to reproduce central life processes, while hunting for emerging phenomena, self-assembly, reaction networks with autocatalytic properties, etc. Furthermore, life is energy-demanding and from a thermodynamic point of view requires constant external drive to maintain its out-of-equilibrium state. However, even minimal chemical systems may exhibit complex energetic transformations, leading to apparently living phenomena. For example, two-component swimmers in water recently demonstrated a novel elastohydrodynamic propulsion mechanism and showcased that even such simple systems can harness energy from temperature fluctuations and store it to recharge [8]. Thus more and more complex processes are moving out of the exclusive domain of biology.

In this chapter, we start with a comparison of the two, in our opinion complementary, syntetic biology approaches, followed by discussion of selected advances in the bottom-up approach. We show how bottom-up synthetic biology combines physical, chemical and biological tools to yield ever more complex artificial structures in a rationally directed and accelerating evolution process, driven by human ingenuity, which may turn out to be a more instructive, definitive, and functional approach than historical speculation. Thereby, our aim is not a comprehensive literature overview but to present a few major developments, revolving around compartmentation, which exemplify the reconstitution and understanding of key living features via minimal synthetic systems made of both natural and man-made constituents. We also attempt to address the limitations and the missing parts in the present blueprint for assembly of an artificial cell as the pieces of a yet undefined puzzle.

The question of origin of life looms large in the last decades and there is optimism among researchers that the problem of synthesizing life in the lab can be solved in the next few decades. Several scientific fields are trying to approach the question from different angles, and synthetic biology has taken a relatively new role in this endeavor. The latter research field has long been dominated by a more top-down approach, modifying existing life and thus developing biotechnological solutions of great practical importance. On the other side, bottom-up synthetic biology – a recently coined and still not firmly established term – shares many of the aims and problems of the origin-of-life research even though in its current iterations it may seem removed from its conceptual ancestor. However, we believe that the ability to build complexity from few components is key to answering "what is life?", and thus to creating life via abiogenesis. All of synthetic biology struggles with the overwhelming complexity of biology, which in turn often seems to doom both approaches' efforts to dead ends. Consequently, for better understanding and control they both resort to the fundamental scientific method of reductionism, too. We illustrate the central features of both concepts in Fig. 1, and compare them in the chapter. The top-down approach always starts with living biological cells and reduces their complexity step by step, while still monitoring principle viability as the decisive feature. It dominates the current perception of the term "synthetic biology" in the biosciences, and also in the society, partly because it has made practical advances in sophisticated genetic engineering methods [9]. The stripping down of genes in that approach to expose the functional skeleton of life, however, has been daunting and not ready yet to serve as a simple chassis on which to install new features. For example, the celebrated genome reduction of *Mycoplasma mycoides* has yielded a "minimal cell" with 473 genes, acknowledging that hundreds (one third) of these are with still unknown function [5]. Thereby, the top-down requirements for success have been somehow more generous since they are defined not by understanding but by apparent biotechnological figures of merit, e.g., the production of a given protein, and not by mechanistic blueprints from molecular and structural biology that serve as incomplete instructions for building a cell. On the contrary, bottom-up synthetic biology usually addresses initially non-living systems with up to 20 chemicals in a bilayer membrane or other physical confinement, where most of the component interactions are known. Remarkably, such dramatically simplified systems have recently been brought to successfully mimic several curiously complex features of living systems. These include motility [8], metabolism [10], communication [11], reproduction [12], and evolution [13], as covered in more detail below. There are also combinations of both top-down and bottom-up approaches, which have intermediate complexity – for instance, ready-to-use protein expression machinery in form of a cell extract, encapsulated in minimal compartment.

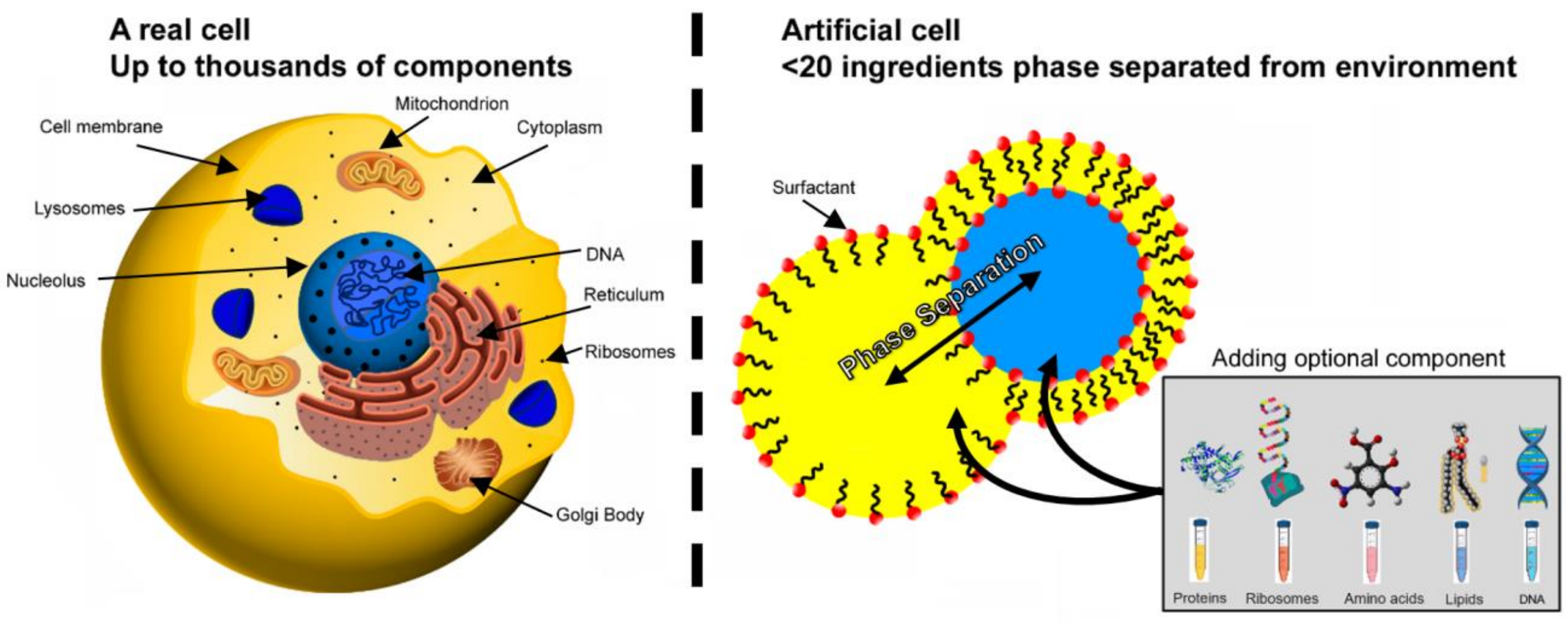

Figure 1. Schematic comparison of minimal biological vs. artificial cells. The most minimal biological cells still have > 1000 components, whereas artificial models have only up to 20, allowing great control and understanding of processes. Of course, artificial cells have yet to become alive – with all the properties expected of life.

For the purpose of engineering features of life, the bottom-up approach within synthetic biology has an unlimited arsenal of building blocks at its disposal. Thus, next to primitive molecules from the speculative toolbox of abiogenesis, highly evolved or engineered protein machinery and even fully synthetic alternatives, products of modern chemistry, are being employed. The bottom-up method aims to pick up prokaryotic and eukaryotic functional units like LEGO pieces, borrows proteins across all kingdoms, tunes and blends, dissembles and reassembles them in minimal or novel ways. In parallel, synthetic chemistry and material science deliver additional means, deliberately or accidentally biomimetic, which may fuel bold speculations about other possible life forms [14,15]. Here, the creative aspect during the quest for the physicochemical principles of life is particularly emphasized, not only as a search for what happened long ago, but how we could use building blocks to start life *de novo*. In other words, studies of molecular assemblies with life-mimicking collective properties should not only help defining life and understanding how it appeared, but will also allow building a living cell, or possibly a tissue, organ or even organism from scratch. Furthermore, ever more complex biological phenomena are becoming addressable in the bottom-up approach, enabled by the evolved molecular toolbox.

Unlike the abiogenesis research, which firmly serves a fundamental role, bottom-up synthetic biology is able to incorporate parallel efforts in colloidal and material science, which also have great application potential. In the long run, once we learn what life is, how to put it together and jump-start it, we may succeed in engineering the perfect platform organism – assembled and tuned to maximize yields and minimize the metabolic penalty for self-preservation, unchained from the muddle of evolutionary heritage. Both top-down and bottom-up approaches are driven in parallel towards this goal. Metaphorically speaking, the top-down biologist pulls cards from a house tower until it falls, while the bottom-up artisan stacks cards until the tower is raised. To tackle this problem, both lines of research in synthetic biology employ a modular tactic, *divide et impera*, which is conceptually reminiscent of conventional mechanical and chemical engineering, hence the parallels to building a car or a chemical plant. In fact, we argue that the modularization of the traditionally fluid, interpenetrating and messy concept of life is one of the hallmarks of the umbrella term synthetic biology. The other defining feature is directly evident from the semantics of "synthetic" and exemplifies the desire to assemble and create at will, analogous to the established field of synthetic chemistry.

Reductionism is a valid and proven method, intrinsic to science, and not exclusive to any research field. Bottom-up synthetic biology tries to reduce biological complexity by creating putative biological entities with certain key functions of life. It can be considered as an evolved form of traditional *in vitro* reconstitution experiments in biochemistry, enabled by continuous technological developments and accumulation of knowledge. We note that abiogenesis is not the only motivation for bottom-up synthetic biology and that such experiments build up the understanding of cells and biology in general. We note in the same way, that top-down synthetic biology has been characterized as nothing but a more powerful genetic engineering, continuously evolving since the first attempts of selection and cross breeding. However, although there may be some truth in synthetic biology being partial rebranding of other fields, we believe that there is increasing momentum and awareness for an impending paradigm shift. Its connection with the famous Feynman quote "*What I cannot create, I do not understand*" will have multiple implications for life sciences and biotechnology. Thus, by creating and understanding new systems, synthetic biology will help reveal the underlying molecular and physical principles of life and will guide the elucidation and discrimination of conflicting abiogenesis hypotheses.

In Table 1, we present a brief summary comparison of the tradeoff made between the top-down and bottom-up approaches in terms of functionality and understanding. Though not alive, the featured bottom-up constructs have drastically fewer, well-characterized components, to which one can reduce the mechanism for the functionality in question and possibly understand the component actions.

Table 1 – Function vs number of components needed

| System type/ Functionality | Top-Down | | Bottom-up | | Representative Photo of bottom-up examples |
|---|---|---|---|---|---|
| | # of chemical components* | Ref | # of chemical components* | Ref | |
| Reproduction | > 1000 | 16 | <10 | 12 | 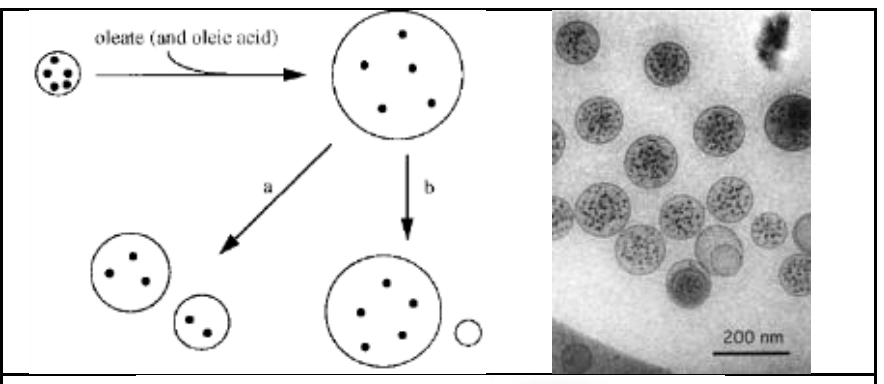 |
| Enclosed metabolism | > 1000 | 16 | 10–50 | 17 | 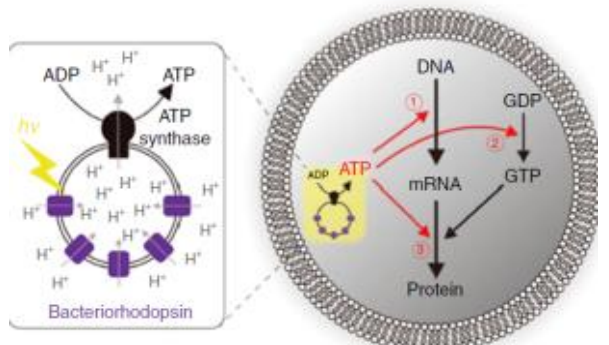 |
| Division/Splitting | > 1000 | 16 | < 20 (vesicle), 2 (oil droplet) | 18, 19 | 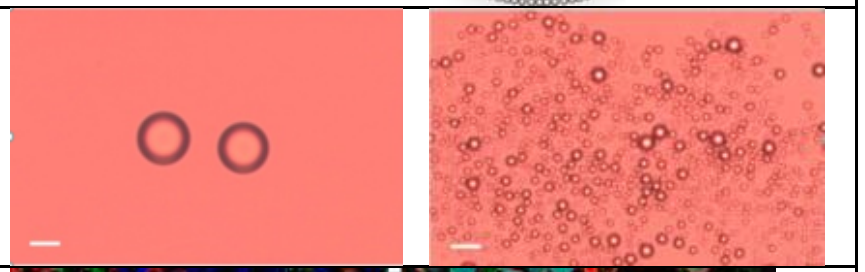 |
| Communications | ?? | ?? | <5–6 | 20 | 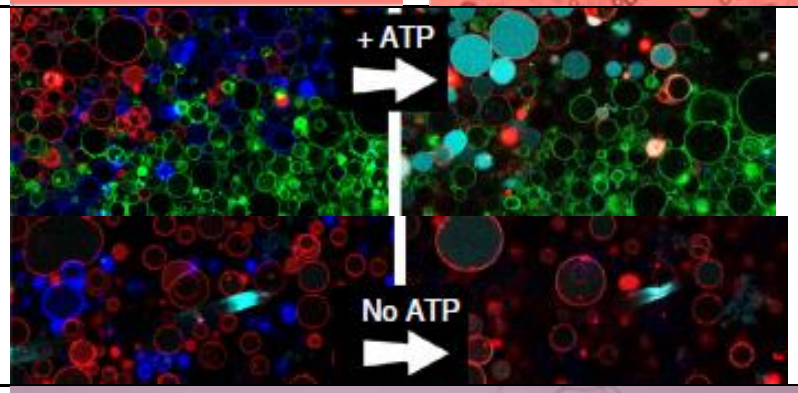 |
| Swimming | ?? | ?? | 2 | 8 | 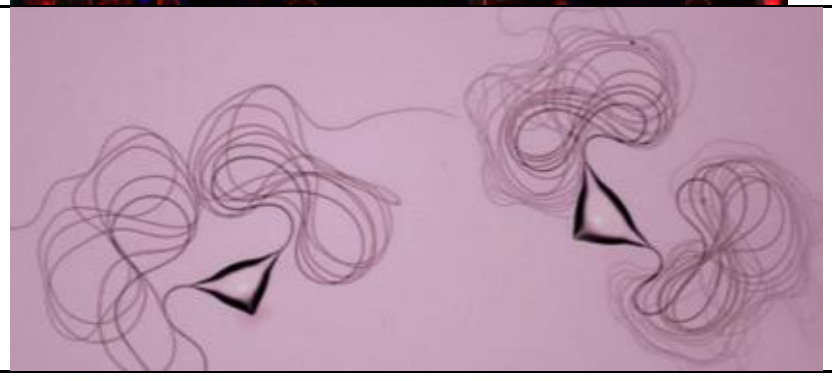 |
| Shape-changing | > 3000 | 21 | 2 | 22 | 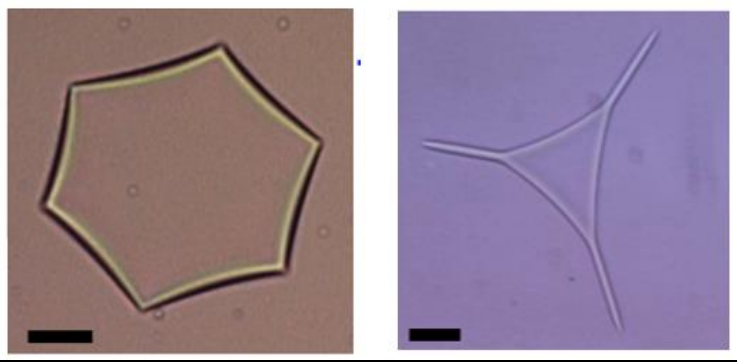 |
| Evolution | > 1000 | 16 | 5 | 23 | 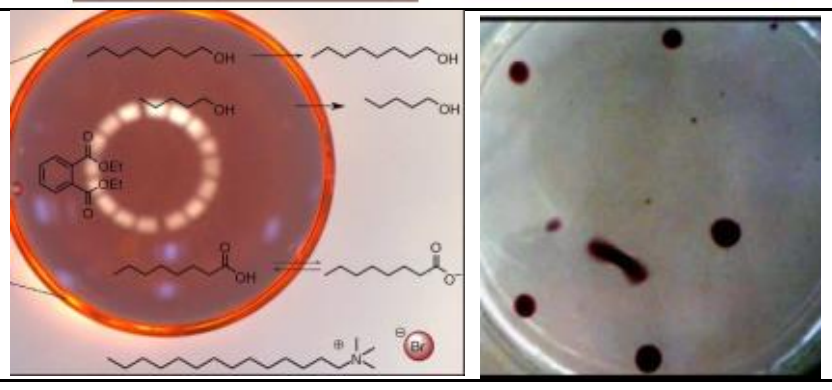 |

* Minimum # of distinct chemical species needed in addition to water

Below we present a concise kaleidoscopic overview of some recent efforts towards the creation of a cell, termed interchangeably proto-, minimal or synthetic, largely depending on the preference of the researchers. We focus on recent progress milestones in compartmentalization, reflecting our own and attempt to relate to other key abiogenesis questions like metabolism and cycles of growth and division. Wefirmly refrain, however, from claims for comprehensiveness, let alone favoring metabolism-first or RNA-first hypotheses.

**Unifying the plausible protocells in line with the crowded cell**

Some form of compartmentation is indispensable for life, even if it should only serve to overcome the dilution of molecules. In fact, compartments do much more – they segregate reactions, enable gradients and bestow individuality, among many other functions. From the simplest protocells, compartments have evolved to today's highly compartmentalized and crowded cells. There are many parallel efforts in trying to decipher their physicochemical and biological origins, and whether membranes came first, or the liquid-liquid phase separation of coacervates. Bottom-up synthetic biology (as a conceptual offspring of the origin-of-life research) has generally not been concerned about the evolutionary timeline and prebiotic plausibility of their building blocks [24]. But by constructing more and more functional models of life, it will show their relative importance for *de novo* origins of life.

Thus, vesicles are being formed from phospholipids, their presumable predecessors fatty acids, and combinations thereof, but also from synthetic molecules such as amphiphilic polymers [25], showcasing the omnipotent molecular principles of self-assembly. These polymersomes, as well as their hybrid combinations with phospholipids, substantiate the chemically synthetic aspect of bottom-up biology, striving to achieve improved properties by deliberate design of building blocks [26,27]. However, they also mimic fundamental phenomena like membrane phase separation [28,29], associated with the raft hypothesis as a means for two-dimensional compartmentation of membrane proteins, which has in turn numerous implications for the organization of life. Different vesicles are now routinely used to accommodate advanced biological processes like mimicking replication via PCR [30] or reverse-transcription PCR [31] and protein synthesis [32].

Alternatively, it may be reasonable to assume that phase separation of an active "proto-cytosol", potentially stabilized by viscosity, has preceded the membrane as a universal boundary, or at least that both variants of compartmentation emerged simultaneously in a synergistic manner. This hypothesis is reinforced by the recent findings and emerging consensus on the ubiquitous role of liquid-liquid phase separation in biology [33]. In parallel to sequestering nanoparticles and biomolecules, and the enhancement of enzyme kinetics [34], coacervates may be coated with fatty acids as a primitive form of a membrane [35]. Such an architecture bridges the apparent dichotomy (largely determined by the preference of the researchers and not so much by conflicting hypotheses) with respect to possible prebiotic compartments, and has been realized with phospholipids and short polyamines [36] but also with fully synthetic polycations [37].

An interesting but still underrepresented aspect is the deliberate control of compartment morphology, as both membraneless and membrane-bound compartments tend to relax to spherical shapes. Thus, the transient non-spherical shape of biomolecular condensates [38] has not been arrested yet, unlike the diatom-reminiscent morphology attainable in hydrocarbon emulsions [22]. Notably, though the exact mechanisms are still under debate and investigation,[39] the causes of shaped archaea may be due to physical phase transitions rather than biological programming [24,40,41]. Control of surface-induced rotator phase transitions has been used for the generation of desired shapes in mixed oil systems [42], bottom-up polymerization of shapes [43], and also triglyceride oils with chain lengths found in the membranes of living cells [44]. Fundamental studies have found that such transformations occur near or above the CMC of the surfactants, ensuring a well-packed layer on the droplet surface.[45] Highly dynamic phase transformations have been harnessed even to replicate biomimetic flagellate-like movement [8]. In parallel, there is a growing number of studies on cytoskeleton reconstitution, which should lead to dynamic morphological asymmetry. However, the shape control is largely still not intrinsic but achieved by microfluidic manipulation of droplets [46] and vesicles [47] instead, to study the assembly of actin and tubulin homologues. Nevertheless, the first successful examples have been already reported – the known bacterial determinant MreB has induced rod-like shapes of PEGylated vesicles [48]. Even though in that case the interactions between cytoskeleton and membrane have been tuned by synthetic means, by consulting the established area of membrane sculpting by proteins and other biomolecules, plausible primitive morphogenesis phase diagrams may be generated in the future. Similarly, non-biological reaction-diffusion processes have been used to transfer patterns from soft to hard materials,[49] or grow nanoscale patterns and three-dimensional structures.[50] The process of artificial morphogenesis has been used to grow prescribed geometric shapes bottom-up – oil droplets transforming due to internal phase transitions – has been shown to apply to a variety of oils [51], some used in biology, which may have been the basis of protobiological shapes. The smallest genome of a modern non-spherical bacterium so far is of the helical Drosophila symbiont *Spiroplasma poulsonii* sHy, with 1584 genes [21], though mapping between genome sizes and shapes of organisms is an open problem that could shed some light on which shapes are more difficult than others.

**Self-sustained cycles of growth and division**

In addition to structures that help sustain it, life is composed of highly dynamic and far out-of-equilibrium processes, which both origin-of-life studies and bottom-up synthetic biology seek to reconstitute. These range from a transient single biochemical reaction [52] to complex interactions and feedback loops enabling oscillatory behavior [53]. Separately, the processes of growth, copying of genetic information, or division have recently been achieved in many contexts. However, the combined process of reproduction of growing entities together with heritable information has been elusive. It is notable that of 7 genes determined required for a normal cell division of the smallest cell so far, only two (ftsZ, sepF) have known function [16]. Thereby, continuous self-reproduction deserves a particular attention, as it is recognized as one of the cornerstones of life and constitutes the basis for autopoiesis, heredity and evolution and thus it is a fruitful area of current inquiry. We provide a short overview of milestones in this direction, as well as other recent perspectives of approaches, suggesting possible solutions of the problem [54,55].

With respect to self-reproduction of protocellular compartments, consecutive cycles of growth and division have been demonstrated predominantly in membrane-bounded compartments. This is also the more challenging case when compared to coacervate growth, where existing frameworks from emulsions (e.g., Ostwald ripening) can be readily applied. The growth of fatty acid vesicles by simple uptake of the same is experimentally attainable thanks to the fast dynamics of these simpler amphiphiles [56]. In some cases, it is succeeded by division [57], while for instance, the temporal coupling with RNA replication [58] would be a matter of synchronization. The realization of primitive strategies for growth of phospholipid vesicles, however, is not as straightforward [59]. In this regard, bottom-up synthetic biology may force the revisiting of mixed approaches, involving both single- and double-chain amphiphiles [60]. Even though the motivation behind such studies might have been different in the past, mixed phospholipid-fatty acid systems bear significant biological resemblance, related to the uptake of exogenous fatty acids and their use for membrane synthesis. In this regard, phospholipid vesicles were shown to grow upon addition of oleic acid, which was then converted to phospholipids by an in vitro assembled eight-enzyme cascade [61]. Similar cascades, encapsulated in liposomes, have been also encoded, translated and even self-regulated via genetic programming of cell-free systems [62]. Cytosol-confined transcription-translation machinery has been also used to grow peptide compartments [63], where the somehow exotic membrane building blocks bear a remote analogy to viral capsid proteins. Polypeptide compartments have been also shown to fuse and used to accommodate DNA ligation [64]. Recursive reproduction of liposomes has been coupled with RNA replication via freeze-thaw cycling over ten iterations [65]. While this approach is not cell-centric but populational, in the sense that the progeny cannot be traced because the reproduction relies on random exchange of membrane and nucleic acids, it holds abiogenetic relevance with respect to the messy conditions at the onset of life. In parallel, membrane growth and division has been realized via synthetic precursors, at the borderline of biology. In one example, the membrane self-reproduction was coupled with DNA amplification [66], where nucleic acids participated in the formation of the catalytic complex, while the combination with fusion also allowed for recursive cycles, referred to as ingestion, replication, maturity and division [67]. In another report, triazole derivatives were used as phospholipid-embedded autocatalysts that drove potentially inexhaustible membrane formation [68]. Synthetic chemistry has been also employed for the formation of polymer membranes, like template polymerization of polyaniline [69] or ring-opening metathesis polymerization coupled to fusion [70], and both approaches remarkably resulted in traceable growth. These examples are obviously unrelated to the origin of life, yet worth mentioning in this context, as they appear to breathe some life into the otherwise inert polymer compartments. From a current standpoint, the fact that most of the enzymes responsible for phospholipid biosynthesis are membrane proteins may appear as a chicken-and-egg problem, which motivates efforts towards *de novo* membrane formation, reenacting the emergence of the first compartments. This has been realized by putative abiogenetic means, such as enzyme-free synthesis of phospholipids [71] but also via engineering of the soluble enzyme FadD10, which is normally responsible for fatty acid activation and coupling with CoA [72].

Growth is also attainable via vesicle fusion, which can proceed in the absence of complex synaptic machinery, driven for instance by simple physicochemical cues like oppositely charged phospholipids [73] or membrane tension by osmotic triggers [74]. Fusion has been mainly investigated in relation to secretory pathways and protein trafficking, and while it could be tentatively put in the context of eukaryotic lipid trafficking, there are growing indications that this is not a major mechanism for plasma membrane expansion [75]. Nevertheless, fusion of protocellular compartments can be still hypothesized as a primitive mechanism for information exchange and integration of functions at the early onset of abiogenesis. Such a phenomenon has been demonstrated via the fusion of complementary gene circuits [76]. These examples, even if they resulted from deliberate manipulation via modern and sophisticated biochemical means, build on evidence for the possibility of the simultaneous and independent emergence of membrane-forming agents on the one side and life-central biopolymers (i.e. nucleic acids and polypeptides) [77] on the other, and reproduce later stages of speculative abiogenetic scenarios. In a purely phenotypical line of thought, spontaneously formed biomolecular condensates could sequester relevant molecules like protopeptides [78] and become decorated with primitive membranes. Such a containment would provide evolutionary advantage via protecting and arresting the fairly dynamic condensates, while transition from fatty acids to phospholipids might have been catalyzed by segregated protoenzymes. If different catalysts are available at the same time, they could chemically “evolve” with the membrane by simple partitioning, following its altered amphiphilic properties, which does not require the central biological dogma of gene expression. Membrane internalization would also turn the protocells fitter with respect to reproduction [79], while potentially catalyzing the polymerization of aminoacids [80] or information carriers [81], alongside other indirect (protective) advantages for the latter [82]. We note that although many of the currently investigated coacervates rely on nucleic acids, other "life-inert" phase-separating molecules [83] may be also sought as potential seeds for life "crystallization".

In an oversimplified analogy to modern cells [84], once the protocell has metabolized and acquired sufficient material, it is ripe to split into two. Thus, next to reconstitution of membrane formation, one of the other current challenges in bottom-up synthetic biology is to reconstitute the machinery for binary fission, even though the precise mechanisms for regulation of the cell size have not been fully elucidated [85]. This bold aim has come into reach by the recently acquired experimental ability to reproducibly place and study cytoskeletal machinery inside vesicles [86]. There has been a significant progress in minimal divisomes, e.g., division of liposomes by bacterial Z rings [87]. However, the assembly and orchestration of several processes [88], including complex reaction networks for positioning and geometry sensing (sizers) [89] or machinery for DNA segregation [90], will require ingenious craftsmanship and will cover the concluding stages of cell shaping. In fact, prebiotic division should have resulted from much simpler physicochemical and intrinsic triggers, and was very likely far from being symmetric. Moreover, modern organisms still exhibit budding, hyphal growth or daughter cell formation.

Rudimental division of liposomes is often accompanied by an imbalance between volume and area and relies on some minimal "cytosol". Such examples have been demonstrated with encapsulated PEG [91] and dextran/PEG, whereby the latter led to asymmetric budding, aided by the phase-separated aqueous system [92]. Biochemically active cytosols also may lead to division – enzymatic hydrolysis of urea enabled the splitting of mixed phospholipid/fatty acid vesicles thanks to the resulting osmotic gradient and susceptibility of oleic acid to pH change [18]. In another case, osmotic deflation was induced by external enzymatic reaction, while the main driving force for parting the liposomes was phase separation of the membrane [93]. While deflation of hollow liposomes has proved as a practical trigger for division and it serves as an indispensable tool when studying membrane physics, it is difficult to ascribe the first division events solely to a floppy, relaxed membrane, provided the simultaneously required nutrient uptake and metabolism in the crowded cytosol. In fact, HeLa cells were found to increase their surface and hydrostatic pressure during mitotic rounding, which might expel water, but the concomitant osmoregulation masked any measurable changes in volume [94].

Other bottom-up division strategies focus on the manipulation by proteins in the framework of membrane remodeling. One such artificial approach employed the adsorption of His-tagged GFP to NTA-functionalized phospholipids at low density and enabled recording morphology diagrams based on the degree of deflation (i.e. the surface-to-volume ratio) and the spontaneous curvature, whereby the curvature also affected the constriction force, necessary for neck fission [95]. This is an example how the theoretical milestones for division become attainable in a minimal system, even though GFP has been externally supplied. The physicochemical fundaments of primitive division can be further solidified by testing other membrane effectors, even if they normally exercise quite different roles in biology. One such example, facilitated by the decreasing cost and wide availability of DNA synthesis, is DNA origami as it currently enables attaining various shapes in a predictable and straightforward way [96]. Thus, DNA origami that mimicked the prominent BAR protein domains resulted in vesicle tubulation for instance [97]. In this regard, computational leaps in protein folding open new horizons for protein design [98], which will hopefully mature to "protein origami" in the near future and will thus revolutionize structural biology. DeepMind's AlphaFold 2 has already achieved prediction of protein structures on par with single crystal experimental methods [99] and now University of Washington researchers have made an open-source framework RoseTTA with similar accuracy available to all [100]. In 2021 the AI structure prediction has deservedly been chosen as Science Magazine discovery of the year, and has enabled not only protein structure prediction but also docking of substrates and protein-protein complex interactions [101]. Among numerous other applications, this synthetic approach will allow identifying minimal and repeating functional sequences in order to draw a family tree back to the first peptides with catalytic or structural functions. The implications of protein folding on the understanding of life and its origin obviously reach far beyond membrane sculpting or cytoskeletal dynamics. Yet, this serves to show the parallel with a readily available technology like DNA origami and exemplify the huge potential in respect to self-reproduction too. Lastly, division relies on mechanical forces, which have been even speculated to precede biochemical energy [102], and therefore only the simultaneous investigation of metabolically and mechanically active membrane and proto-cytosol will lead to more conclusive theories. From a thermodynamic point of view, living processes have been thought necessary to split droplets and vesicles in a controlled way (rather than via mechanical shear). Yet, recently oil droplets subjected to temperature fluctuations have harnessed energy of rotator phase transitions and split spontaneously into higher energy (smaller diameter) droplets [19]. Molecular dynamics methods have only been designed this year to try and capture the molecular details of such solid-solid rotator phase transitions [103]. This and other chemical ratchets are among the many abiotic strategies that still wait to be discovered in the area of reproduction and may be potentially transferrable to systems comprising today's or ancestral building blocks of life.

At the same time, other paradigmatic patterns are being assembled from scratch too. On the level of proteins, the reconstitution of the cell-division-related bacterial MinCDE system has yielded insights into how chemical energy, such as ATP, can be harnessed to promote spatiotemporal protein self-organization, resulting in the emergence of micron-scale pattern formation [104,105], directional protein transport [106] and even the mechanical transformation of membranes [107]. On the level of transcription-translation, a minimal large-genome replicator, variation of the PURE system, was also able to regenerate its proteins for translation and transcription [108], while a similar construct was chemostatically operated in a microfluidic device to sustain the activity [109]. Moreover, the ability of the former system for regeneration of essential proteins was tested in serial transfer experiments [110], which conditions mimicked cellular reproduction in a few generations. Furthermore, the autocatalytic replication of a Φ25 bacteriophage genome was performed successfully, encapsulated in liposomes [111]. Integrating these studies in a thought experiment, along with the potentially attainable recursive self-reproduction of compartments that was discussed above, substantiates the hope for experimental verification of minimal life, which may in turn streamline the search for genesis.

**Transport and energy generation at the interface**

Compartmentation necessitates transport in and out to enable the continuous processing of matter, analogous to the microfluidic chemostat just mentioned above. In coacervates, this would be simply driven by the interplay of chemical equilibria and partitioning effects, but in membrane-bounded compartments, there is a barrier to be overcome. It can be anticipated that this transport was a stochastic process at the early onset of life, when various environmental or intrinsic triggers randomly caused membrane defects. Defects can also be artificially and reliably induced by electric fields in electroporation, a useful current technique for studying membrane biophysics [112]. It is also a practical tool for vesicle loading, e.g., to produce protein and genetic-material packed nano-vesicles (exosomes) which are useful in gene-therapy and an essential vehicle in cellular communication [113]. Interestingly, in recent experiments, while the loading efficiency of DNA in the nano-sized exosomes depended on the DNA size in the latter case, there was no difference in translocation (loading) efficiency in the range of 25–20,000 bp for micron giant unilamellar vesicles [114], implying a size-dependent difference in mechanism. A natural pore opening is not a guarantee for prebiotic genetic translocation, as it may occur on very long timescales and potentially lead to rupture. However, the presence of some mechanical reinforcement on the inside could maintain compartment integrity, as demonstrated by the resilience of liposomes loaded with artificial DNA-cytoskeleton against osmotic shocks [115]. Provided that some degree of integrity is ensured, membrane defects could be reproduced by means that are more abiogenetically relevant – for instance by UV exposure that would degrade phospholipids. In this regard, transient permeability was speculated as a mechanism for DNA exchange between liposomes subjected to freeze-thaw cycles, as opposed to secretory fission and fusion [116].

***Energy and complexity*** There is a dichotomy between the smooth running metabolic engine of biology and understanding how one could design it. Its complexity is one of the mysteries of life, no less because it did not emerge fully formed either, but evolved over time to what we see today. Though many life forms use similar reactions in their metabolism, not one life form can do them all. We live in an ecosystem that life itself has built over billions of years. Until the second world war, virtually all (90–95%) of the nitrogen life used was being produced by no more than 13 phyla of bacteria, capable of making nitrogenases and most of them relying rare metals, either Mo or V [117]. The Haber-Bosch process has enabled intensive agriculture and in recent years has almost matched the volume of Nature's fixing of nitrogen [118]. Yet the industrial catalysts can only perform the reaction above 400 °C and at high pressures of 50–200 bar, whereas nitrogenases can perform the reaction at room temperature and ambient pressure around 1 bar [119]. The long-standing mystery of nitrogenases has been unraveled in the last few years [117] and parallel efforts have now allowed us to design small molecule catalysts that also perform nitrogen fixation at room temperature [120]. But we are still far away from understanding the complex symbioses that the nitrogenase-synthesizing bacteria have formed with the roots of legumes and a select few other plants. The main point we would like to exemplify here is about complexity. Besides the extreme complexity of life, there could be major, simple (even if less efficient) milestones that we can create bottom-up. Thus the bottom-up creation of a minimal nitrogen-fixing system would not only create an industrially desirable catalyst but if integrated in a simple metabolic cycle, it would also help create a simplified version of current biology, perhaps not unlike the simple proto-metabolic cycles present at the origin of life.

Complexities across the metabolism have made truly astounding transformations possible. Glycolysis and its associated electron transport chain can split the energy of a glucose molecule into 30–32 molecules of ATP. This raises the intriguing question, is there a fundamentally smallest unit of energy that life can harvest? Is that environmentally determined, and how many kT of energy is it? In modern living systems this capability is spectacular, it has been tuned down to about 3 kT in some circumstances, modern cells living on just formate as a source of energy. These cells are able to access chunks of energy smaller than the barrier of translocating a single ion across their membranes, by coupling opposing ion pumps in complex structures called antiporter proteins [121,122]. Then, it is possible for cells to subsist on energy spikes that are barely large enough to stick out of the thermal background motion. But did these capabilities emerge together with the first life forms? Unlikely. The high functionality in biology is accompanied by complexity that we are barely starting to understand, and one that certainly did not come overnight. The electron transport chain alone incorporates over 60 associated proteins that would have evolved from many less efficient processes. Even small parts of it, e.g. the ATP Synthase enzyme have extremely complex structures to have emerged at once, and would have needed simpler functional precursors. In the absence of finely tuned coupled processes, a key technique used by evolution was likely again said compartmentation.

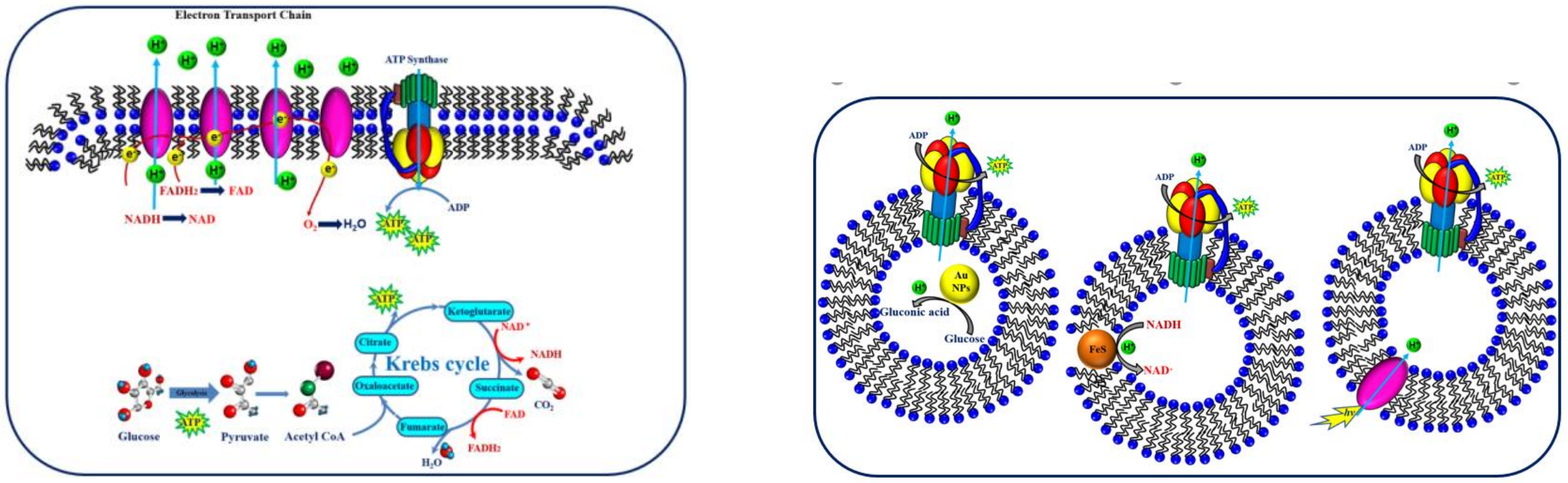

Figure 2. Comparing metabolic complexity in living cells and attempts to engineer artificial cells. The many dozens of coupled reactions and proteins in current living cells could not emerge at the same time. The much simpler to understand and control systems attempted in synthetic biology are also closer to a potential origin of life, which later evolved to the kind of complexity we see today.

***Energy compartmentation*** The evolution of selective protein pores is apparently difficult to be traced back to the early timeline of life, which in fact holds true for all proteins, but todays' engineering capabilities [123] may allow for identification of minimal motifs, aided by combinatorial screening of peptide libraries [124]. Combinatorial testing can be also applied with respect to the membrane, in order to determine key mechanical properties that influence protein folding, such as bending rigidity [125]. Synthetic chemistry can turn out to be useful for this endeavor, because it potentially allows obtaining more degrees of design freedom – an exemplary approach linked the surface pressure of different polymer/phospholipid mixtures to the distribution of potassium channels [126]. Even though the motivation behind such studies is application-driven, and these polymers have no biological relation whatsoever, the concomitant generation of data will lead to a better understanding of the required membrane properties, which can be then extrapolated to the early membranes.

The reconstitution of unspecific pores like hemolysin has proved useful in multiple instances of assembled biomimetic constructs. Cell-free expression of the toxin inside liposomes enabled prolonged synthesis of GFP due to the enhanced uptake of nutrients [127], while in another case it facilitated the export of glucose as a chemical signal for communication with proteinosomes [11]. Reconstituted hemolysin has been also used to activate intracellular signaling thanks to Ca permeation and mechanosensitive channels [128]. The preference for this relatively simple protein pore is largely due to its facile insertion in membranes, which has even allowed for evolution of hemolysin mutants via liposome display from encapsulated translation machinery [129]. Yet it unequivocally demonstrates the importance of permeability mechanisms, after the eventual sealing of the early membranes, alongside the fact that many processes can be entertained without high specificity. Upon conceptual merging of the above examples, one could easily imagine liposome-encapsulated cell-free expression of simple peptides for screening of their pore-forming properties, whereby compartment hits that ensure more efficient nutrient uptake would have evolutionary advantage.

In the field of energy metabolism, bottom-up constructs also generate complementary knowledge. From an origin-of-life standpoint, the stabilization of FeS clusters with peptides resulted in primitive catalysts that could reproduce the functions of complex I with respect to NADH oxidation, electron transfer to quinones, and change of pH arising from the redox reaction, which provided a plausible roadmap for the chemical origin of energetic machinery [10]. These prebiotic catalysts essentially mimicked the functions of non-pumping (type 2) NADH dehydrogenases, which are the single entry point for electrons in several organisms, while proton translocation in vesicles can be accomplished by redox shuttles alone [130]. Interestingly, shuttles with fully synthetic spiropyran chemistry have even enabled the light-driven generation of transmembrane proton gradients [131] to mock rhodopsins, which suggests that proton pumps might emerge from relatively simple chemistry. In this regard, light was also used for NADH oxidation by a polymeric photocatalyst through the generation of superoxide [132] (accompanying also the natural oxidation by complex I as a side reaction [133]), while subsequent coupling with a silica-encapsulated enzyme system for cellular defense counteracted the oxidative stress [134]. This synthetic example is difficult to be conceptually linked to possible cellular energetics, as the electrons from NADH would be wasted in vain. However, it motivates an abiotic revisiting of the reverse reaction, in line with the reported photocatalytic activity of $TiO_2$ [135] and Au [136] nanoparticles towards $NAD^+$ reduction, and synthesis of nucleoside bases [137], within the search for primitive and decoupled pathways for carbon and "hydrogen fixation".

On the other hand, the experimental modularity and tunability of a fully functional, yet minimal construct for respiration and oxidative phosphorylation (referred to as artificial mitochondrion, comprising complex I, oxidase and ATP synthase embedded in phospholipid vesicles [138]) might be employed towards better mechanistic understanding. For instance, it has been suggested that protons generated in the final step of respiration traverse along the membrane, as evidenced by the effect of lateral distance between *bo₃* oxidase and ATP synthase on the ATP synthesis rates [139]. This partially conflicted the textbook cartoon of cytoplasm acidification with a direct "kinetic" coupling instead, and underlined the importance of protein clustering [140]. The association of protons with phospholipids is known and has been already experimentally reflected in the altered buffering capacity of liposome suspensions. These findings have implications for the overall architecture needed for oxidative phosphorylation and may turn obsolete the putative need of periplasmic (or mitochondrial intermembrane) space as proton storage in the context of protocells. In other words, just the interface by itself might have been sufficient at the beginning. Moreover, the proton gradient may be stoichiometrically generated from multiple other reactions, as shown for instance by the integration of gold nanoparticle-catalyzed glucose oxidation with ATPase [141]. Similar minimal assemblies like the Arc cascade, that included an arginine antiporter, have been built to reproduce substrate-level phosphorylation too [142]. However, the gap between purely mineral [143] or chemical scenarios [144] and modern ATP synthases and kinases has not been bridged yet by constructs of intermediate complexity, although studies in the field of enzyme mimics can prove helpful in this regard [145]. The protein apparatus for light- [146] and chemically [147] driven ATP synthesis has been also reconstituted in polymer vesicles. Apart from a purely creative rationale, such demonstrations once again help to determine the necessary membrane properties (thickness, fluidity, softness) to accommodate protein machinery [148], which in turn should limit the chemical landscape of potential membrane-forming molecules. Photosynthetic ATP production via bacteriorhodopsin and ATPase has been also addressed in combination with cell-free expression in giant liposomes, which provided a mechanism for partial replenishment of the bioenergetic apparatus [17].

Origin-of-life studies (extensive discussion in [149]) and bottom-up synthetic biology tackle the energy metabolism from two different directions, while the necessity of compartments, regardless of their particular nature, and the segregation of redox chemistry to the interface is recognized in both fields. Yet the experimental proficiency and modular assembly, highlighted in the bottom-up approach, may be employed to shed further light on abiogenesis. This will require expansion of the state-of-the-art biochemical toolbox towards plausible early energetic scenarios by consulting and reproducing cryptic metabolisms and not clinging to the currently prevalent ones. In this line of thought, the complexity of the rotor-stator-type ATPase conflicts its evolutionary conservation, while oxygen, which might not have been widely available, is not a must even for eukaryotic mitochondria [150], let alone for the numerous microbes relying on iron respiration [151]. On a practical note, oxygen depletion is in fact a common hurdle in multiple in vitro experiments. Moreover, the nearly ubiquitous phosphate bond, epitomized in the universality of ATP for energy storage, can be confronted with the possibility of phosphate-free metabolism [152]. Thus, a synthetic rationale to redirect the assembly of synthetic cells to underexplored simpler and ancient energy pathways may prove especially useful for the abiogenesis conundrum, while potentially providing unexpected utilities.

**Synergistic effects towards the origin of life**

Two conceptual scenarios crystalize from the above bottom-up studies: either an entirely biological (modern-day) toolbox, yet minimized and reorganized, or on the other hand combinations of biologically evolved and potential primitive parts from a presumed abiotic genesis. Both approaches contribute to the understanding of life by the identification of minimal assemblies that reenact biological functions. Furthermore, some works engage entirely synthetic building blocks. These man-made materials may have had no abiogenetic aspirations, however they also aid the elucidation physicochemical principles underlying overarching "biological" processes and patterns. With increasing sophistication synthetic chemistry nowadays is creating nucleic (DNA, RNA) sequences, peptides of increasing length and sophistication, phospholipid-like amphiphiles, as well as many other molecules associated previously only with the domain of life. Judicious interpenetration has resulted in artificial muscle polymers with additional functions,[153] including even self-sensing materials.[154] New developments, such as sequence-defined polymers [155] promise peptide-like functionality from non-peptide chemistry, and perhaps if life is created *de novo* in the lab, it may augment rather than just recreate existing life. The ambitious aim of abiogenic creation of life by straightforward experimental verification, in contrast to the more speculative nature of historical origin-of-life research, promises to bring such origins in the realm of modern synthetic biology [6]. In line with its engineering aptitude, bottom-up synthetic biology often resorts to technological platforms for compartmentation, such as droplet-based microfluidics for the production of vesicles and coacervates or cell-free systems for protein expression. These are useful tools for studying later snapshots of abiogenesis, after life is seeded via compartmentation, and allow for addressing multiple evolutionary effects. However, the experimental reenactment of the true transition from chemistry to biology will require integrated efforts. As many of the fundamental principles of that transition are still poorly understood, a partial withdrawal from deliberate manipulation to achieve life, towards combinatorial and autonomous approaches, may allow greater chances of discovery and success.

We conclude with a conceptual comparison between a popular literature reference – Frankenstein's monster [156] – and the anthropomorphic *golem* as a recursively appearing theme in mythology and religion. The latter creature is made of clay and somewhat amorphous, clearly echoing the abiogenic approach to life. While its animation is a subject of wide interpretation, in many of the recorded legends golems have been made to do useful hard work, also mirroring motivations in biotechnology and bottom-up robotic design.[157,158] On the contrary, Mary Shelley's fictional character was assembled from human parts, bolted together by artificial materials, and brought to life by electricity. Thus, we can metaphorically summarize the contribution of bottom-up synthetic biology to the origin of life by using these parallels, whereby we note that the prevalently negative connotation of Frankenstein's creature is a close-minded rejection based on superficial unfamiliarity and bias, but it is, as a matter of fact, a human being, capable of emotions. Thus, the ongoing fabrication of Frankenstein's creatures, whose limbs are occasionally replaced by clay mockups or biomechatronic prosthetics, may help us discover how Adam was created from mud.

References:

1 Miller, S. L. A Production of Amino Acids Under Possible Primitive Earth Conditions. *Science* **117**, 528-529 (1953).

2 Miller, S. L. & Urey, H. C. Organic Compound Synthesis on the Primitive Earth. *Science* **130**, 245-251, doi:http://doi.org/10.1126/science.130.3370.245 (1959).

3 Sharma, S., Arya, A., Cruz, R. & Cleaves II, H. J. Automated Exploration of Prebiotic Chemical Reaction Space: Progress and Perspectives. *Life (Basel)* **11**, 1140, doi:http://doi.org/10.3390/life11111140 (2021).

4 Wołos, A. *et al.* Synthetic connectivity, emergence, and self-regeneration in the network of prebiotic chemistry. *Science* **369**, eaaw1955, doi:http://doi.org/10.1126/science.aaw1955 (2020).

5 Hutchison, C. A. *et al.* Design and synthesis of a minimal bacterial genome. *Science* **351**, aad6253, doi:http://doi.org/10.1126/science.aad6253 (2016).

6 Schwille, P. *et al.* MaxSynBio: Avenues Towards Creating Cells from the Bottom Up. *Angew Chem Int Ed Engl* **57**, 13382-13392, doi:http://doi.org/10.1002/anie.201802288 (2018).

7 Hesiod. *The Theogony*. (https://www.gutenberg.org/files/348/348-h/348-h.htm#chap26, c. 730-700 B.C.).

8 Cholakova, D. *et al.* Rechargeable self-assembled droplet microswimmers driven by surface phase transitions. *Nature Physics* **17**, 1050-1055, doi:http://doi.org/10.1038/s41567-021-01291-3 (2021).

9 Benner, S. A. & Sismour, A. M. Synthetic biology. *Nature Reviews Genetics* **6**, 533-543, doi:http://doi.org/10.1038/nrg1637 (2005).

10 Bonfio, C. *et al.* Prebiotic iron–sulfur peptide catalysts generate a pH gradient across model membranes of late protocells. *Nature Catalysis* **1**, 616-623, doi:http://doi.org/10.1038/s41929-018-0116-3 (2018).
11 Tang, T. Y. D. *et al.* Gene-Mediated Chemical Communication in Synthetic Protocell Communities. *ACS Synthetic Biology* **7**, 339-346, doi:http://doi.org/10.1021/acssynbio.7b00306 (2018).
12 Berclaz, N., Müller, M., Walde, P. & Luisi, P. L. Growth and Transformation of Vesicles Studied by Ferritin Labeling and Cryotransmission Electron Microscopy. *The Journal of Physical Chemistry B* **105**, 1056-1064, doi:http://doi.org/10.1021/jp001298i (2001).
13 Decraene, J., Mitchell, G. G. & McMullin, B. in *Advances in Biologically Inspired Information Systems Studies in Computational Intelligence* (eds Dressler F. & Carreras I.) 165-184 (Springer, 2007).
14 Bains, W. Many Chemistries Could Be Used to Build Living Systems. *Astrobiology* **4**, 137-167, doi:http://doi.org/10.1089/153110704323175124 (2004).
15 Kan, S. B. J., Lewis, R. D., Chen, K. & Arnold, F. H. Directed evolution of cytochrome c for carbon–silicon bond formation: Bringing silicon to life. *Science* **354**, 1048, doi:http://doi.org/10.1126/science.aah6219 (2016).
16 Pelletier, J. F. *et al.* Genetic requirements for cell division in a genomically minimal cell. *Cell* **184**, 2430-2440 e2416, doi:http://doi.org/10.1016/j.cell.2021.03.008 (2021).
17 Berhanu, S., Ueda, T. & Kuruma, Y. Artificial photosynthetic cell producing energy for protein synthesis. *Nature Communications* **10**, 1325, doi:http://doi.org/10.1038/s41467-019-09147-4 (2019).
18 Miele, Y. *et al.* Self-division of giant vesicles driven by an internal enzymatic reaction. *Chemical Science* **11**, 3228-3235, doi:http://doi.org/10.1039/C9SC05195C (2020).
19 Tcholakova, S. *et al.* Efficient self-emulsification via cooling-heating cycles. *Nat Commun* **8**, 15012, doi:http://doi.org/10.1038/ncomms15012 (2017).
20 Buddingh, B. C., Elzinga, J. & van Hest, J. C. M. Intercellular communication between artificial cells by allosteric amplification of a molecular signal. *Nat Commun* **11**, 1652, doi:http://doi.org/10.1038/s41467-020-15482-8 (2020).
21 Gerth, M. *et al.* Rapid molecular evolution of Spiroplasma symbionts of Drosophila. *Microb Genom* **7**, doi:http://doi.org/10.1099/mgen.0.000503 (2021).
22 Denkov, N., Tcholakova, S., Lesov, I., Cholakova, D. & Smoukov, S. K. Self-shaping of oil droplets via the formation of intermediate rotator phases upon cooling. *Nature* **528**, 392-395, doi:http://doi.org/10.1038/nature16189 (2015).
23 Gutierrez, J. M., Hinkley, T., Taylor, J. W., Yanev, K. & Cronin, L. Evolution of oil droplets in a chemorobotic platform. *Nat Commun* **5**, 5571, doi:http://doi.org/10.1038/ncomms6571 (2014).
24 van hest, J. C. M. in *COLF: Conflicting Models of the Origin of Life* (ed S. K. Smoukov, Gordon, R., Seckbach, J.) (Wiley-Scrivener Publishing, 2022).
25 Discher, B. M. *et al.* Polymersomes: Tough Vesicles Made from Diblock Copolymers. *Science* **284**, 1143, doi:http://doi.org/10.1126/science.284.5417.1143 (1999).
26 Rideau, E., Dimova, R., Schwille, P., Wurm, F. R. & Landfester, K. Liposomes and polymersomes: a comparative review towards cell mimicking. *Chemical Society Reviews* **47**, 8572-8610, doi:http://doi.org/10.1039/C8CS00162F (2018).
27 Petit, J. *et al.* A modular approach for multifunctional polymersomes with controlled adhesive properties. *Soft Matter* **14**, 894-900, doi:http://doi.org/10.1039/C7SM01885A (2018).
28 Dao, T. P. T. *et al.* Phase Separation and Nanodomain Formation in Hybrid Polymer/Lipid Vesicles. *ACS Macro Letters* **4**, 182-186, doi:http://doi.org/10.1021/mz500748f (2015).
29 Rideau, E., Wurm, F. R. & Landfester, K. Membrane Engineering: Phase Separation in Polymeric Giant Vesicles. *Small* **16**, 1905230, doi:https://doi.org/10.1002/smll.201905230 (2020).

30 Sato, Y., Komiya, K., Kawamata, I., Murata, S. & Nomura, S.-i. M. Isothermal amplification of specific DNA molecules inside giant unilamellar vesicles. *Chemical Communications* **55**, 9084-9087, doi:http://doi.org/10.1039/C9CC03277K (2019).
31 Tsugane, M. & Suzuki, H. Reverse Transcription Polymerase Chain Reaction in Giant Unilamellar Vesicles. *Scientific Reports* **8**, 9214, doi:http://doi.org/10.1038/s41598-018-27547-2 (2018).
32 Nishimura, K. *et al.* Cell-Free Protein Synthesis inside Giant Unilamellar Vesicles Analyzed by Flow Cytometry. *Langmuir* **28**, 8426-8432, doi:10.1021/la3001703 (2012).
33 Hyman, A. A., Weber, C. A. & Jülicher, F. Liquid-Liquid Phase Separation in Biology. *Annual Review of Cell and Developmental Biology* **30**, 39-58, doi:http://doi.org/10.1146/annurev-cellbio-100913-013325 (2014).
34 Koga, S., Williams, D. S., Perriman, A. W. & Mann, S. Peptide–nucleotide microdroplets as a step towards a membrane-free protocell model. *Nature Chemistry* **3**, 720-724, doi:http://doi.org/10.1038/nchem.1110 (2011).
35 Dora Tang, T. Y. *et al.* Fatty acid membrane assembly on coacervate microdroplets as a step towards a hybrid protocell model. *Nature Chemistry* **6**, 527-533, doi:http://doi.org/10.1038/nchem.1921 (2014).
36 Aumiller, W. M., Pir Cakmak, F., Davis, B. W. & Keating, C. D. RNA-Based Coacervates as a Model for Membraneless Organelles: Formation, Properties, and Interfacial Liposome Assembly. *Langmuir* **32**, 10042-10053, doi:http://doi.org/10.1021/acs.langmuir.6b02499 (2016).
37 Pir Cakmak, F., Grigas, A. T. & Keating, C. D. Lipid Vesicle-Coated Complex Coacervates. *Langmuir* **35**, 7830-7840, doi:http://doi.org/10.1021/acs.langmuir.9b00213 (2019).
38 Spoelstra, W. K., van der Sluis, E. O., Dogterom, M. & Reese, L. Nonspherical Coacervate Shapes in an Enzyme-Driven Active System. *Langmuir* **36**, 1956-1964, doi:http://doi.org/10.1021/acs.langmuir.9b02719 (2020).
39 Haas, P. A., Goldstein, R. E., Cholakova, D., Denkov, N. & Smoukov, S. K. Comment on "Faceting and Flattening of Emulsion Droplets: A Mechanical Model". *Phys Rev Lett* **126**, 259801, doi:http://doi.org/10.1103/PhysRevLett.126.259801 (2021).
40 Gordon, R., Hanczyc, M. M., Denkov, N. D., Tiffany, M. A. & Smoukov, S. K. in *Habitability of the Universe Before Earth* (eds Richard Gordon & Alexei A. Sharov) 427-490 (Academic Press, 2018).
41 Gordon, R. in *COLF: Conflicting Models of the Origin of Life* (ed S. K. Smoukov, Gordon, R., Seckbach, J.) (Wiley-Scrivener Publishing, 2022).
42 Cholakova, D., Valkova, Z., Tcholakova, S., Denkov, N. & Smoukov, S. K. "Self-Shaping" of Multicomponent Drops. *Langmuir* **33**, 5696-5706, doi:http://doi.org/10.1021/acs.langmuir.7b01153 (2017).
43 Lesov, I. *et al.* Bottom-Up Synthesis of Polymeric Micro- and Nanoparticles with Regular Anisotropic Shapes. *Macromolecules* **51**, 7456-7462, doi:http://doi.org/10.1021/acs.macromol.8b00529 (2018).
44 Yeagle, P. L. in *The Membranes of Cells (Third Edition)* (ed Philip L. Yeagle) 27-56 (Academic Press, 2016).
45 Feng, J. *et al.* Minimum Surfactant Concentration Required for Inducing Self-shaping of Oil Droplets and Competitive Adsorption Effects. *Soft Matter*, accepted (2022).
46 Fanalista, F. *et al.* Shape and Size Control of Artificial Cells for Bottom-Up Biology. *ACS Nano* **13**, 5439-5450, doi:http://doi.org/10.1021/acsnano.9b00220 (2019).
47 Ganzinger, K. A. *et al.* FtsZ Reorganization Facilitates Deformation of Giant Vesicles in Microfluidic Traps**. *Angewandte Chemie International Edition* **59**, 21372-21376, doi:https://doi.org/10.1002/anie.202001928 (2020).

48 Garenne, D., Libchaber, A. & Noireaux, V. Membrane molecular crowding enhances MreB polymerization to shape synthetic cells from spheres to rods. *Proceedings of the National Academy of Sciences* **117**, 1902, doi:http://doi.org/10.1073/pnas.1914656117 (2020).
49 Smoukov, S. K., Bishop, K. J. M., Klajn, R., Campbell, C. J. & Grzybowski, B. A. Cutting into solids with micropatterned gels. *Advanced Materials* **17**, 1361-1365, doi:http://doi.org/10.1002/adma.200402086 (2005).
50 Smoukov, S. K., Bitner, A., Campbell, C. J., Kandere-Grzybowska, K. & Grzybowski, B. A. Nano- and microscopic surface wrinkles of linearly increasing heights prepared by periodic precipitation. *Journal of the American Chemical Society* **127**, 17803-17807, doi:http://doi.org/10.1021/ja054882j (2005).
51 Cholakova, D., Denkov, N., Tcholakova, S., Lesov, I. & Smoukov, S. K. Control of drop shape transformations in cooled emulsions. *Adv Colloid Interface Sci* **235**, 90-107, doi:http://doi.org/10.1016/j.cis.2016.06.002 (2016).
52 Beneyton, T. *et al.* Out-of-equilibrium microcompartments for the bottom-up integration of metabolic functions. *Nature Communications* **9**, 2391, doi:http://doi.org/10.1038/s41467-018-04825-1 (2018).
53 Semenov, S. N. *et al.* Rational design of functional and tunable oscillating enzymatic networks. *Nature Chemistry* **7**, 160-165, doi:http://doi.org/10.1038/nchem.2142 (2015).
54 Olivi, L. *et al.* Towards a synthetic cell cycle. *Nat Commun* **12**, 4531, doi:http://doi.org/10.1038/s41467-021-24772-8 (2021).
55 Wang, C., Yang, J. & Lu, Y. Modularize and Unite: Toward Creating a Functional Artificial Cell. *Front Mol Biosci* **8**, 781986, doi:http://doi.org/10.3389/fmolb.2021.781986 (2021).
56 Chen, I. A. & Szostak, J. W. A Kinetic Study of the Growth of Fatty Acid Vesicles. *Biophysical Journal* **87**, 988-998, doi:https://doi.org/10.1529/biophysj.104.039875 (2004).
57 Zhu, T. F. & Szostak, J. W. Coupled Growth and Division of Model Protocell Membranes. *Journal of the American Chemical Society* **131**, 5705-5713, doi:http://doi.org/10.1021/ja900919c (2009).
58 O'Flaherty, D. K. *et al.* Copying of Mixed-Sequence RNA Templates inside Model Protocells. *Journal of the American Chemical Society* **140**, 5171-5178, doi:http://doi.org/10.1021/jacs.8b00639 (2018).
59 Ivanov, I. *et al.* Directed Growth of Biomimetic Microcompartments. *Advanced Biosystems* **3**, 1800314, doi:https://doi.org/10.1002/adbi.201800314 (2019).
60 Peterlin, P., Arrigler, V., Kogej, K., Svetina, S. & Walde, P. Growth and shape transformations of giant phospholipid vesicles upon interaction with an aqueous oleic acid suspension. *Chemistry and Physics of Lipids* **159**, 67-76, doi:https://doi.org/10.1016/j.chemphyslip.2009.03.005 (2009).
61 Exterkate, M., Caforio, A., Stuart, M. C. A. & Driessen, A. J. M. Growing Membranes In Vitro by Continuous Phospholipid Biosynthesis from Free Fatty Acids. *ACS Synthetic Biology* **7**, 153-165, doi:http://doi.org/10.1021/acssynbio.7b00265 (2018).
62 Blanken, D., Foschepoth, D., Serrão, A. C. & Danelon, C. Genetically controlled membrane synthesis in liposomes. *Nature Communications* **11**, 4317, doi:http://doi.org/10.1038/s41467-020-17863-5 (2020).
63 Frank, T., Vogele, K., Dupin, A., Simmel, F. C. & Pirzer, T. Growth of Giant Peptide Vesicles Driven by Compartmentalized Transcription–Translation Activity. *Chemistry – A European Journal* **26**, 17356-17360, doi:https://doi.org/10.1002/chem.202003366 (2020).
64 Schreiber, A., Huber, M. C. & Schiller, S. M. Prebiotic Protocell Model Based on Dynamic Protein Membranes Accommodating Anabolic Reactions. *Langmuir* **35**, 9593-9610, doi:10.1021/acs.langmuir.9b00445 (2019).

65 Tsuji, G., Fujii, S., Sunami, T. & Yomo, T. Sustainable proliferation of liposomes compatible with inner RNA replication. *Proceedings of the National Academy of Sciences*, 201516893, doi:http://doi.org/10.1073/pnas.1516893113 (2015).

66 Kurihara, K. *et al.* Self-reproduction of supramolecular giant vesicles combined with the amplification of encapsulated DNA. *Nature Chemistry* **3**, 775-781, doi:http://doi.org/10.1038/nchem.1127 (2011).

67 Kurihara, K. *et al.* A recursive vesicle-based model protocell with a primitive model cell cycle. *Nature Communications* **6**, 8352, doi:http://doi.org/10.1038/ncomms9352 (2015).

68 Hardy, M. D. *et al.* Self-reproducing catalyst drives repeated phospholipid synthesis and membrane growth. *Proceedings of the National Academy of Sciences* **112**, 8187, doi:http://doi.org/10.1073/pnas.1506704112 (2015).

69 Kurisu, M. *et al.* Reproduction of vesicles coupled with a vesicle surface-confined enzymatic polymerisation. *Communications Chemistry* **2**, 117, doi:http://doi.org/10.1038/s42004-019-0218-0 (2019).

70 Varlas, S. *et al.* Polymerization-Induced Polymersome Fusion. *Journal of the American Chemical Society* **141**, 20234-20248, doi:http://doi.org/10.1021/jacs.9b10152 (2019).

71 Liu, L. *et al.* Enzyme-free synthesis of natural phospholipids in water. *Nature Chemistry* **12**, 1029-1034, doi:http://doi.org/10.1038/s41557-020-00559-0 (2020).

72 Bhattacharya, A., Brea, R. J., Niederholtmeyer, H. & Devaraj, N. K. A minimal biochemical route towards de novo formation of synthetic phospholipid membranes. *Nature Communications* **10**, 300, doi:http://doi.org/10.1038/s41467-018-08174-x (2019).

73 Lira, R. B., Robinson, T., Dimova, R. & Riske, K. A. Highly Efficient Protein-free Membrane Fusion: A Giant Vesicle Study. *Biophysical Journal* **116**, 79-91, doi:https://doi.org/10.1016/j.bpj.2018.11.3128 (2019).

74 Deshpande, S., Wunnava, S., Hueting, D. & Dekker, C. Membrane Tension–Mediated Growth of Liposomes. *Small* **15**, 1902898, doi:https://doi.org/10.1002/smll.201902898 (2019).

75 Voelker, D. R. Bridging gaps in phospholipid transport. *Trends in Biochemical Sciences* **30**, 396-404, doi:https://doi.org/10.1016/j.tibs.2005.05.008 (2005).

76 Adamala, K. P., Martin-Alarcon, D. A., Guthrie-Honea, K. R. & Boyden, E. S. Engineering genetic circuit interactions within and between synthetic minimal cells. *Nature Chemistry* **9**, 431-439, doi:http://doi.org/10.1038/nchem.2644 (2017).

77 Kitadai, N. & Maruyama, S. Origins of building blocks of life: A review. *Geoscience Frontiers* **9**, 1117-1153, doi:https://doi.org/10.1016/j.gsf.2017.07.007 (2018).

78 Frenkel-Pinter, M., Samanta, M., Ashkenasy, G. & Leman, L. J. Prebiotic Peptides: Molecular Hubs in the Origin of Life. *Chemical Reviews* **120**, 4707-4765, doi:http://doi.org/10.1021/acs.chemrev.9b00664 (2020).

79 Adamala, K. P., Engelhart, A. E. & Szostak, J. W. Collaboration between primitive cell membranes and soluble catalysts. *Nature Communications* **7**, 11041, doi:http://doi.org/10.1038/ncomms11041 (2016).

80 Fukuda, K., Shibasaki, Y., Nakahara, H. & Liu, M.-h. Spontaneous formation of polypeptides in the interfacial thin films of amphiphilic amino acid esters: acceleration of the polycondensation and control of the structure of resultant polymers. *Advances in Colloid and Interface Science* **87**, 113-145, doi:https://doi.org/10.1016/S0001-8686(99)00041-X (2000).

81 Rajamani, S. *et al.* Lipid-assisted Synthesis of RNA-like Polymers from Mononucleotides. *Origins of Life and Evolution of Biospheres* **38**, 57-74, doi:http://doi.org/10.1007/s11084-007-9113-2 (2008).

82 Adamala, K. & Szostak, J. W. Nonenzymatic Template-Directed RNA Synthesis Inside Model Protocells. *Science* **342**, 1098, doi:http://doi.org/10.1126/science.1241888 (2013).

83 Chandru, K. *et al.* Simple prebiotic synthesis of high diversity dynamic combinatorial polyester libraries. *Communications Chemistry* **1**, 30, doi:http://doi.org/10.1038/s42004-018-0031-1 (2018).
84 Cai, L. & Tu, B. P. Driving the Cell Cycle Through Metabolism. *Annual Review of Cell and Developmental Biology* **28**, 59-87, doi:http://doi.org/10.1146/annurev-cellbio-092910-154010 (2012).
85 Cadart, C., Venkova, L., Recho, P., Lagomarsino, M. C. & Piel, M. The physics of cell-size regulation across timescales. *Nature Physics* **15**, 993-1004, doi:10.1038/s41567-019-0629-y (2019).
86 Weiss, M. *et al.* Sequential bottom-up assembly of mechanically stabilized synthetic cells by microfluidics. *Nature Materials* **17**, 89-96, doi:http://doi.org/10.1038/nmat5005 (2018).
87 Osawa, M. & Erickson, H. P. Liposome division by a simple bacterial division machinery. *Proceedings of the National Academy of Sciences* **110**, 11000, doi:http://doi.org/10.1073/pnas.1222254110 (2013).
88 Hürtgen, D., Härtel, T., Murray, S. M., Sourjik, V. & Schwille, P. Functional Modules of Minimal Cell Division for Synthetic Biology. *Advanced Biosystems* **3**, 1800315, doi:https://doi.org/10.1002/adbi.201800315 (2019).
89 Schweizer, J. *et al.* Geometry sensing by self-organized protein patterns. *Proceedings of the National Academy of Sciences* **109**, 15283, doi:http://doi.org/10.1073/pnas.1206953109 (2012).
90 Hürtgen, D. *et al.* Reconstitution and Coupling of DNA Replication and Segregation in a Biomimetic System. *ChemBioChem* **20**, 2633-2642, doi:https://doi.org/10.1002/cbic.201900299 (2019).
91 Terasawa, H., Nishimura, K., Suzuki, H., Matsuura, T. & Yomo, T. Coupling of the fusion and budding of giant phospholipid vesicles containing macromolecules. *Proceedings of the National Academy of Sciences*, doi:http://doi.org/10.1073/pnas.1120327109 (2012).
92 Andes-Koback, M. & Keating, C. D. Complete Budding and Asymmetric Division of Primitive Model Cells To Produce Daughter Vesicles with Different Interior and Membrane Compositions. *Journal of the American Chemical Society* **133**, 9545-9555, doi:http://doi.org/10.1021/ja202406v (2011).
93 Dreher, Y., Jahnke, K., Bobkova, E., Spatz, J. P. & Göpfrich, K. Division and regrowth of phase-separated giant unilamellar vesicles. *Angewandte Chemie International Edition* **n/a**, doi:https://doi.org/10.1002/anie.202014174 (2020).
94 Fischer-Friedrich, E., Hyman, A. A., Jülicher, F., Müller, D. J. & Helenius, J. Quantification of surface tension and internal pressure generated by single mitotic cells. *Scientific Reports* **4**, 6213, doi:http://doi.org/10.1038/srep06213 (2014).
95 Steinkühler, J. *et al.* Controlled division of cell-sized vesicles by low densities of membrane-bound proteins. *Nature Communications* **11**, 905, doi:http://doi.org/10.1038/s41467-020-14696-0 (2020).
96 Rothemund, P. W. K. Folding DNA to create nanoscale shapes and patterns. *Nature* **440**, 297-302, doi:http://doi.org/10.1038/nature04586 (2006).
97 Franquelim, H. G., Khmelinskaia, A., Sobczak, J.-P., Dietz, H. & Schwille, P. Membrane sculpting by curved DNA origami scaffolds. *Nature Communications* **9**, 811, doi:http://doi.org/10.1038/s41467-018-03198-9 (2018).
98 Service, R. F. 'The game has changed.' AI triumphs at protein folding. *Science* **370**, 1144, doi:http://doi.org/10.1126/science.370.6521.1144 (2020).
99 Jumper, J. *et al.* Highly accurate protein structure prediction with AlphaFold. *Nature* **596**, 583-589, doi:http://doi.org/10.1038/s41586-021-03819-2 (2021).
100 Baek, M. *et al.* Accurate prediction of protein structures and interactions using a three-track neural network. *Science* **373**, 871-876, doi:http://doi.org/10.1126/science.abj8754 (2021).
101 Humphreys Ian, R. *et al.* Computed structures of core eukaryotic protein complexes. *Science* **374**, eabm4805, doi:http://doi.org/10.1126/science.abm4805.

102 Hansma, H. G. Mechanical Energy before Chemical Energy at the Origins of Life? *Sci* **2**, doi:http://doi.org/10.3390/sci2040088 (2020).
103 Burrows, S. A., Korotkin, I., Smoukov, S. K., Boek, E. & Karabasov, S. Benchmarking of Molecular Dynamics Force Fields for Solid-Liquid and Solid-Solid Phase Transitions in Alkanes. *J Phys Chem B* **125**, 5145-5159, doi:http://doi.org/10.1021/acs.jpcb.0c07587 (2021).
104 Kretschmer, S., Ganzinger, K. A., Franquelim, H. G. & Schwille, P. Synthetic cell division via membrane-transforming molecular assemblies. *BMC Biology* **17**, 43, doi:http://doi.org/10.1186/s12915-019-0665-1 (2019).
105 Loose, M., Fischer-Friedrich, E., Ries, J., Kruse, K. & Schwille, P. Spatial Regulators for Bacterial Cell Division Self-Organize into Surface Waves in Vitro. *Science* **320**, 789, doi:http://doi.org/10.1126/science.1154413 (2008).
106 Ramm, B. *et al.* The MinDE system is a generic spatial cue for membrane protein distribution in vitro. *Nature Communications* **9**, 3942, doi:http://doi.org/10.1038/s41467-018-06310-1 (2018).
107 Litschel, T., Ramm, B., Maas, R., Heymann, M. & Schwille, P. Beating Vesicles: Encapsulated Protein Oscillations Cause Dynamic Membrane Deformations. *Angewandte Chemie International Edition* **57**, 16286-16290, doi:https://doi.org/10.1002/anie.201808750 (2018).
108 Libicher, K., Hornberger, R., Heymann, M. & Mutschler, H. In vitro self-replication and multicistronic expression of large synthetic genomes. *Nature Communications* **11**, 904, doi:http://doi.org/10.1038/s41467-020-14694-2 (2020).
109 Lavickova, B., Laohakunakorn, N. & Maerkl, S. J. A partially self-regenerating synthetic cell. *Nature Communications* **11**, 6340, doi:http://doi.org/10.1038/s41467-020-20180-6 (2020).
110 Libicher, K. & Mutschler, H. Probing self-regeneration of essential protein factors required for in vitro translation activity by serial transfer. *Chemical Communications* **56**, 15426-15429, doi:http://doi.org/10.1039/D0CC06515C (2020).
111 van Nies, P. *et al.* Self-replication of DNA by its encoded proteins in liposome-based synthetic cells. *Nature Communications* **9**, 1583, doi:http://doi.org/10.1038/s41467-018-03926-1 (2018).
112 Riske, K. A. & Dimova, R. Electro-Deformation and Poration of Giant Vesicles Viewed with High Temporal Resolution. *Biophysical Journal* **88**, 1143-1155, doi:https://doi.org/10.1529/biophysj.104.050310 (2005).
113 Lamichhane, T. N., Raiker, R. S. & Jay, S. M. Exogenous DNA Loading into Extracellular Vesicles via Electroporation is Size-Dependent and Enables Limited Gene Delivery. *Molecular Pharmaceutics* **12**, 3650-3657, doi:http://doi.org/10.1021/acs.molpharmaceut.5b00364 (2015).
114 Sachdev, S. *et al.* DNA translocation to giant unilamellar vesicles during electroporation is independent of DNA size. *Soft Matter* **15**, 9187-9194, doi:http://doi.org/10.1039/C9SM01274E (2019).
115 Kurokawa, C. *et al.* DNA cytoskeleton for stabilizing artificial cells. *Proceedings of the National Academy of Sciences*, 201702208, doi:http://doi.org/10.1073/pnas.1702208114 (2017).
116 Litschel, T. *et al.* Freeze-thaw cycles induce content exchange between cell-sized lipid vesicles. *New Journal of Physics* **20**, doi:http://doi.org/10.1088/1367-2630/aabb96 (2018).
117 Seefeldt, L. C. *et al.* Reduction of Substrates by Nitrogenases. *Chemical Reviews* **120**, 5082-5106, doi:http://doi.org/10.1021/acs.chemrev.9b00556 (2020).
118 Lehnert, N., Musselman, B. W. & Seefeldt, L. C. Grand challenges in the nitrogen cycle. *Chem Soc Rev* **50**, 3640-3646, doi:http://doi.org/10.1039/d0cs00923g (2021).
119 Vojvodic, A. *et al.* Exploring the limits: A low-pressure, low-temperature Haber–Bosch process. *Chemical Physics Letters* **598**, 108-112, doi:http://doi.org/10.1016/j.cplett.2014.03.003 (2014).
120 Ashida, Y. & Nishibayashi, Y. Catalytic conversion of nitrogen molecule into ammonia using molybdenum complexes under ambient reaction conditions. *Chemical Communications* **57**, 1176-1189, doi:http://doi.org/10.1039/D0CC07146C (2021).

121 Kim, Y. J. *et al.* Formate-driven growth coupled with H(2) production. *Nature* **467**, 352-355, doi:http://doi.org/10.1038/nature09375 (2010).
122 Lim, J. K., Mayer, F., Kang, S. G. & Muller, V. Energy conservation by oxidation of formate to carbon dioxide and hydrogen via a sodium ion current in a hyperthermophilic archaeon. *Proc Natl Acad Sci U S A* **111**, 11497-11502, doi:http://doi.org/10.1073/pnas.1407056111 (2014).
123 Ayub, M. & Bayley, H. Engineered transmembrane pores. *Current Opinion in Chemical Biology* **34**, 117-126, doi:https://doi.org/10.1016/j.cbpa.2016.08.005 (2016).
124 Krauson, A. J., He, J., Wimley, A. W., Hoffmann, A. R. & Wimley, W. C. Synthetic Molecular Evolution of Pore-Forming Peptides by Iterative Combinatorial Library Screening. *ACS Chemical Biology* **8**, 823-831, doi:http://doi.org/10.1021/cb300598k (2013).
125 Booth, P. J. *et al.* Evidence That Bilayer Bending Rigidity Affects Membrane Protein Folding. *Biochemistry* **36**, 197-203, doi:10.1021/bi962200m (1997).
126 Kowal, J., Wu, D., Mikhalevich, V., Palivan, C. G. & Meier, W. Hybrid Polymer–Lipid Films as Platforms for Directed Membrane Protein Insertion. *Langmuir* **31**, 4868-4877, doi:http://doi.org/10.1021/acs.langmuir.5b00388 (2015).
127 Noireaux, V. & Libchaber, A. A vesicle bioreactor as a step toward an artificial cell assembly. *Proceedings of the National Academy of Sciences of the United States of America* **101**, 17669, doi:http://doi.org/10.1073/pnas.0408236101 (2004).
128 Hindley, J. W. *et al.* Building a synthetic mechanosensitive signaling pathway in compartmentalized artificial cells. *Proceedings of the National Academy of Sciences* **116**, 16711, doi:http://doi.org/10.1073/pnas.1903500116 (2019).
129 Fujii, S., Matsuura, T., Sunami, T., Kazuta, Y. & Yomo, T. In vitro evolution of α-hemolysin using a liposome display. *Proceedings of the National Academy of Sciences*, 201314585, doi:http://doi.org/10.1073/pnas.1314585110 (2013).
130 Milshteyn, D., Cooper, G. & Deamer, D. Chemiosmotic energy for primitive cellular life: Proton gradients are generated across lipid membranes by redox reactions coupled to meteoritic quinones. *Scientific Reports* **9**, 12447, doi:http://doi.org/10.1038/s41598-019-48328-5 (2019).
131 Xie, X., Crespo, G. A., Mistlberger, G. & Bakker, E. Photocurrent generation based on a light-driven proton pump in an artificial liquid membrane. *Nature Chemistry* **6**, 202-207, doi:http://doi.org/10.1038/nchem.1858 (2014).
132 Ma, B. C. *et al.* Polymer-Based Module for NAD+ Regeneration with Visible Light. *ChemBioChem* **20**, 2593-2596, doi:https://doi.org/10.1002/cbic.201900093 (2019).
133 Kussmaul, L. & Hirst, J. The mechanism of superoxide production by NADH:ubiquinone oxidoreductase (complex I) from bovine heart mitochondria. *Proceedings of the National Academy of Sciences* **103**, 7607, doi:http://doi.org/10.1073/pnas.0510977103 (2006).
134 Jo, S.-M., Zhang, K. A. I., Wurm, F. R. & Landfester, K. Mimic of the Cellular Antioxidant Defense System for a Sustainable Regeneration of Nicotinamide Adenine Dinucleotide (NAD). *ACS Applied Materials & Interfaces* **12**, 25625-25632, doi:http://doi.org/10.1021/acsami.0c05588 (2020).
135 Chen, D., Yang, D., Wang, Q. & Jiang, Z. Effects of Boron Doping on Photocatalytic Activity and Microstructure of Titanium Dioxide Nanoparticles. *Industrial & Engineering Chemistry Research* **45**, 4110-4116, doi:http://doi.org/10.1021/ie0600902 (2006).
136 Roy, S., Jain, V., Kashyap, R. K., Rao, A. & Pillai, P. P. Electrostatically Driven Multielectron Transfer for the Photocatalytic Regeneration of Nicotinamide Cofactor. *ACS Catalysis* **10**, 5522-5528, doi:http://doi.org/10.1021/acscatal.0c01478 (2020).
137 Senanayake, S. D. & Idriss, H. Photocatalysis and the origin of life: Synthesis of nucleoside bases from formamide on TiO<sub>2</sub>(001) single surfaces. *Proceedings of the National Academy of Sciences* **103**, 1194, doi:http://doi.org/10.1073/pnas.0505768103 (2006).

138 Biner, O., Fedor, J. G., Yin, Z. & Hirst, J. Bottom-Up Construction of a Minimal System for Cellular Respiration and Energy Regeneration. *ACS Synthetic Biology* **9**, 1450-1459, doi:http://doi.org/10.1021/acssynbio.0c00110 (2020).
139 Sjöholm, J. *et al.* The lateral distance between a proton pump and ATP synthase determines the ATP-synthesis rate. *Scientific Reports* **7**, 2926, doi:http://doi.org/10.1038/s41598-017-02836-4 (2017).
140 Toth, A. *et al.* Kinetic coupling of the respiratory chain with ATP synthase, but not proton gradients, drives ATP production in cristae membranes. *Proceedings of the National Academy of Sciences* **117**, 2412, doi:http://doi.org/10.1073/pnas.1917968117 (2020).
141 Xu, Y. *et al.* Nanozyme-Catalyzed Cascade Reactions for Mitochondria-Mimicking Oxidative Phosphorylation. *Angewandte Chemie International Edition* **58**, 5572-5576, doi:https://doi.org/10.1002/anie.201813771 (2019).
142 Pols, T. *et al.* A synthetic metabolic network for physicochemical homeostasis. *Nature Communications* **10**, 4239, doi:http://doi.org/10.1038/s41467-019-12287-2 (2019).
143 Gull, M. *et al.* Nucleoside phosphorylation by the mineral schreibersite. *Scientific Reports* **5**, 17198, doi:http://doi.org/10.1038/srep17198 (2015).
144 Gibard, C., Bhowmik, S., Karki, M., Kim, E.-K. & Krishnamurthy, R. Phosphorylation, oligomerization and self-assembly in water under potential prebiotic conditions. *Nature Chemistry* **10**, 212-217, doi:http://doi.org/10.1038/nchem.2878 (2018).
145 Mertes, M. P. & Mertes, K. B. Polyammonium macrocycles as catalysts for phosphoryl transfer: the evolution of an enzyme mimic. *Accounts of Chemical Research* **23**, 413-418, doi:10.1021/ar00180a003 (1990).
146 Choi, H.-J. & Montemagno, C. D. Artificial Organelle: ATP Synthesis from Cellular Mimetic Polymersomes. *Nano Letters* **5**, 2538-2542, doi:http://doi.org/10.1021/nl051896e (2005).
147 Otrin, L. *et al.* Toward Artificial Mitochondrion: Mimicking Oxidative Phosphorylation in Polymer and Hybrid Membranes. *Nano Letters* **17**, 6816-6821, doi:http://doi.org/10.1021/acs.nanolett.7b03093 (2017).
148 Marušič, N. *et al.* Constructing artificial respiratory chain in polymer compartments: Insights into the interplay between bo3 oxidase and the membrane. *Proceedings of the National Academy of Sciences* **117**, 15006, doi:http://doi.org/10.1073/pnas.1919306117 (2020).
149 Sousa, F. L. *et al.* Early bioenergetic evolution. *Philosophical Transactions of the Royal Society B: Biological Sciences* **368**, 20130088, doi:http://doi.org/10.1098/rstb.2013.0088 (2013).
150 Tielens, A. G. M., Rotte, C., van Hellemond, J. J. & Martin, W. Mitochondria as we don't know them. *Trends in Biochemical Sciences* **27**, 564-572, doi:https://doi.org/10.1016/S0968-0004(02)02193-X (2002).
151 Bird, L. J., Bonnefoy, V. & Newman, D. K. Bioenergetic challenges of microbial iron metabolisms. *Trends in Microbiology* **19**, 330-340, doi:https://doi.org/10.1016/j.tim.2011.05.001 (2011).
152 Goldford, J. E., Hartman, H., Smith, T. F. & Segrè, D. Remnants of an Ancient Metabolism without Phosphate. *Cell* **168**, 1126-1134.e1129, doi:http://doi.org/10.1016/j.cell.2017.02.001 (2017).
153 Khaldi, A., Plesse, C., Vidal, F. & Smoukov, S. K. Smarter Actuator Design with Complementary and Synergetic Functions. *Adv Mater* **27**, 4418-4422, doi:http://doi.org/10.1002/adma.201500209 (2015).
154 Wang, T. *et al.* Electroactive polymers for sensing. *Interface Focus* **6**, 20160026, doi:http://doi.org/10.1098/rsfs.2016.0026 (2016).
155 Lutz, J.-F. *Sequence-Controlled Polymers*. (Wiley-VCH, 2018).
156 Shelley, M. *Frankenstein; or, the Modern Prometheus*. (https://www.gutenberg.org/ebooks/42324, 1818).

157 Smoukov, S. K. Sustainably Grown: The Underdog Robots of the Future **in** Mazzolai et al. Roadmap on Soft Robotics, *Multifunct. Mater.* . *arxiv:2206.10306*, doi:http://doi.org/10.48550/arXiv.2206.10306 (2022).
158 Mazzolai, B. & al., e. Roadmap on soft robotics: multifunctionality, adaptability and growth without borders. *Multifunct. Mater.* **in press**, doi:http://doi.org/10.1088/2399-7532/ac4c95 (2022).